\documentclass[%
aip,
amsmath,amssymb,
preprint,
]{revtex4-2}

\usepackage{graphicx}
\usepackage{dcolumn}
\usepackage{bm}

\usepackage[utf8]{inputenc}
\usepackage[T1]{fontenc}
\usepackage{mathptmx}
\usepackage{etoolbox}

\makeatletter
\def\@email#1#2{%
\endgroup
\patchcmd{\titleblock@produce}
{\frontmatter@RRAPformat}
{\frontmatter@RRAPformat{\produce@RRAP{*#1\href{mailto:#2}{#2}}}\frontmatter@RRAPformat}
{}{}
}%
\makeatother
\begin{document}


\title{Voltage-Dependent Photocurrent and Crossover Behavior in Crystalline Silicon Solar Cells}
\author{Toru Matsuura}
\author{Satoshi Saijo}%
\affiliation{ 
Electrical and Electronic Engineering Division, National Institute of Technology (KOSEN), Fukui College, Geshi-cho, Sabae, Fukui 916-8507, Japan
}%

\date{\today}

\begin{abstract}
Contrary to the conventional assumption of voltage-independent photocurrent, we show that the photocurrent in monocrystalline silicon p-n junction solar cells depends explicitly on voltage. 
By applying physically realistic boundary conditions to the classical p-n junction solar-cell model, we derive an analytical expression for the photocurrent that incorporates both the built-in potential and the applied voltage. 
The model predicts a crossover point between the dark and illuminated $I-V$ characteristics, irrespective of illumination intensity. 
This prediction is experimentally validated by lock-in measurements on a commercial monocrystalline silicon solar cell.
These results establish a more consistent understanding of the quantitative role of the built-in potential in the photovoltaic effect.
\end{abstract}



\maketitle

\section{Introduction}
Strong voltage dependence of the photocurrent has been observed and extensively discussed in amorphous silicon \cite{Crandall1982JAP,Merten1998IEEE}, organic \cite{Limpinsel2010PRB,Gaur2013PinPhoto,Kumar2013JAP,Manda2018IEEE,Amorim2020PRApplied}, thin-film \cite{Hegedus2007ProgPV,Klahr2011APL,Moore2014IEEE,Ramesh2022npj}, and perovskite solar cells \cite{Sandberg2020Adv}, where charge separation is strongly governed by the built-in potential, or equivalently by the electric field in the absorber layer.
As a result of the voltage dependence of the photocurrent, these solar cells exhibit a crossover point at which the illuminated and dark $I-V$  characteristics intersect.
 
Paradoxically, for crystalline silicon p-n junction solar cells, which are among the oldest and simplest photovoltaic devices, the photocurrent has long been assumed to be voltage independent. 
Conventional theories of photovoltaic behavior in p-n junctions have largely neglected the quantitative influence of the built-in potential on photocurrent generation.
Instead, voltage-dependent features observed in experimental current-voltage characteristics have often been attributed to parasitic series and shunt resistances, rather than to a fundamental voltage dependence of the photocurrent itself.
Relatively few studies have investigated the voltage dependence of the photocurrent in crystalline silicon solar cells \cite{Robinson1994JAP,Hishikawa2000JJAP}.

Since charge separation is governed by the built-in potential, the photocurrent would be expected to exhibit some voltage dependence.
However, most textbooks treat the photocurrent generated in a p-n junction as a constant in analytical formulations \cite{Grove1967text,Yang1978text,Sze2001text,Anderson2005text,Wurfel2009text}, as expressed by
\begin{eqnarray}
I_{ph} = -qA G (L_n + L_p),
\label{I_ph}
\end{eqnarray}
or 
\begin{eqnarray}
I_{ph} = -qA G (L_n +W+ L_p),
\label{I_phW}
\end{eqnarray}
where $q$ is the elementary charge, and $A$ is the area of the p-n junction.
$L_n$ and $L_p$ are the diffusion lengths of minority carriers, where the subscripts $n$ and $p$ refer to electrons in the p-type semiconductor and holes in the n-type semiconductor, respectively.
$W$ is the depletion-layer width.
$G$ represents the net generation rate of electron-hole pairs per unit volume induced by externally incident photons.
In these expressions, $I_{ph}$ is assumed to be independent of both the applied voltage and the built-in potential.
The constant-photocurrent model has generally been regarded as sufficient to explain the behavior of crystalline silicon p-n junction solar cells.
However, it remains unclear whether the photocurrent is truly independent of the applied voltage.



\begin{figure}[t]
\centering
\includegraphics[width=0.7\linewidth]{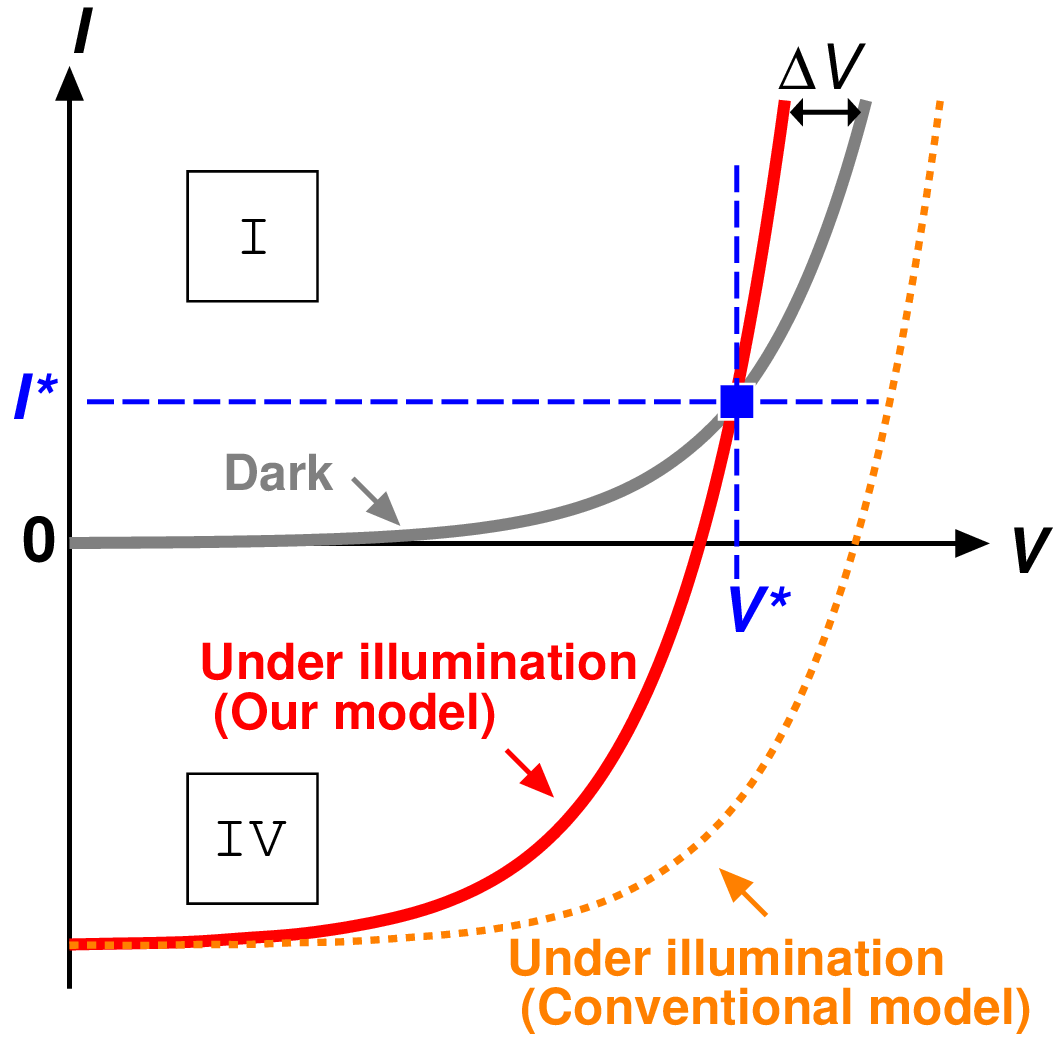} 
\vspace{8cm}
\caption{The $I-V$ characteristics of a p-n junction under dark and illuminated conditions, based on the conventional model [Eq. (\ref{I_ph})] and our model [Eq. (\ref{I_ph2})], are presented. The crossover point at ($V^*$, $I^*$) in the first quadrant [I] is indicated. }
\label{fig1}
\end{figure}

In this manuscript, we revisit the conventional model of p–n junction solar cells and derive a simple analytical expression for the photocurrent which depends on the built-in potential $V_{bi}$, and the applied voltage $V$, given by
\begin{eqnarray}
I_{ph} = -qA G(L_n + L_p)\left\{1 - \alpha \exp\left(-\frac{q(V_{bi} - V)}{k_B T}\right)\right\},
\label{I_ph2}
\end{eqnarray}
where $k_B$ is the Boltzmann constant, and $T$ is the temperature.
$\alpha$ is a positive constant determined by the semiconductor parameters.
Equation (\ref{I_ph2}) predicts that even conventional solar cells, such as crystalline silicon solar cells, exhibit voltage-dependent photocurrent when $V$ approaches $V_{bi}$, leading to crossover behavior in the first quadrant of the $I-V$ characteristics, as shown in Fig. \ref{fig1}.
We experimentally verify the predicted crossover behavior in commercial monocrystalline silicon solar cells under illumination using an optical lock-in measurement technique. 
The crossover behavior is not due to thermally induced shifts in the  $I-V$ characteristics. 
The observed crossover behavior provides direct evidence that the photocurrent in monocrystalline silicon solar cells is not constant but depends on both the applied voltage and the built-in potential.






\section{Theory}
\begin{figure}[t]
\centering
\includegraphics[width=0.7\linewidth]{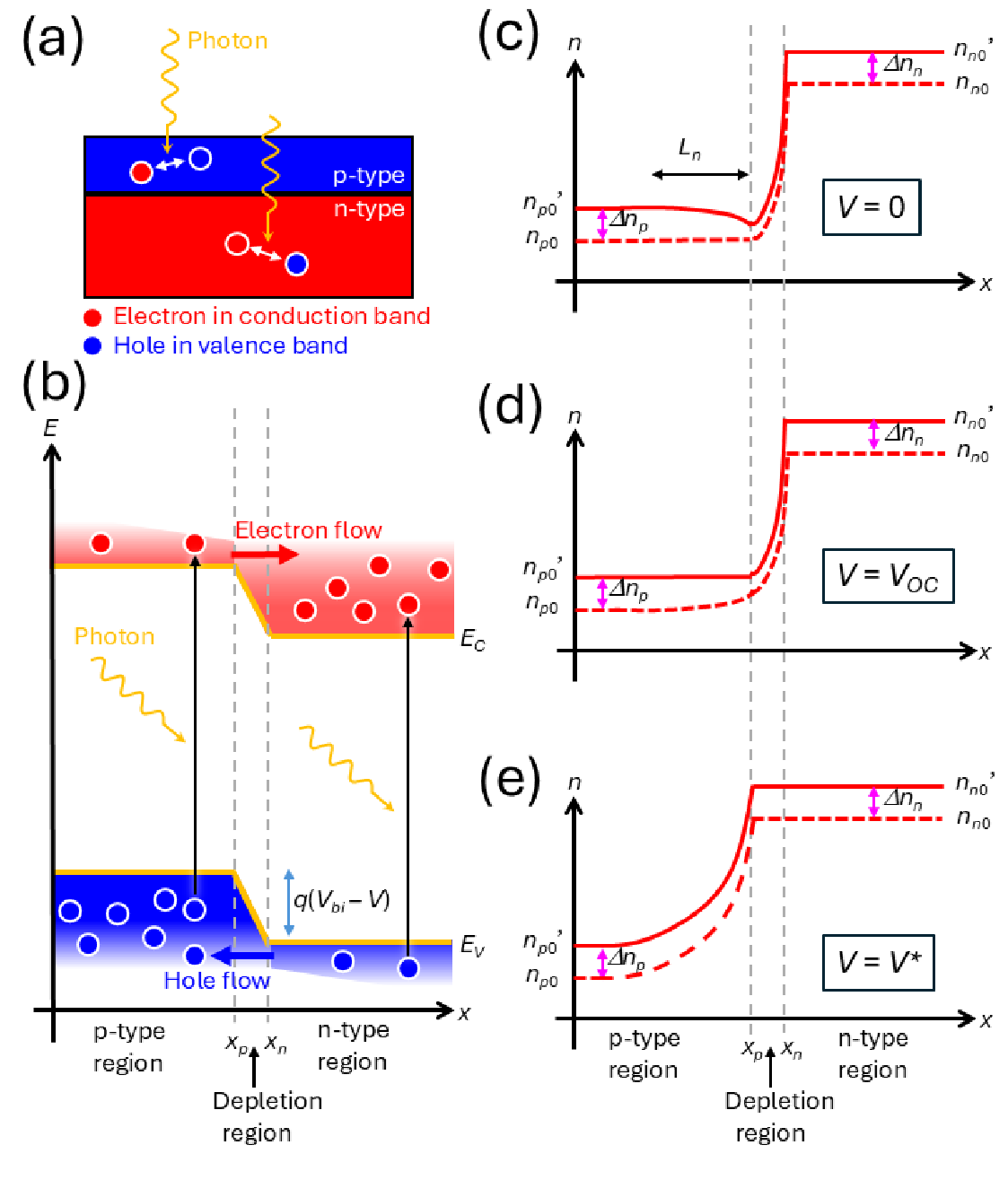} 
\vspace{0cm}
\caption{(a) Photoexcitation in a p-n junction solar cell under illumination.
(b) Band diagram of a p-n junction under illumination under illumination ($G > 0$). 
$E_C$ and $E_V$ denote the conduction and valence band-edge energies, respectively.
$V_{bi}$ is the built-in potential, and $V$ is the voltage between the cathode and anode, and it varies depending on the current. 
(c) - (e) Schematic illustration of the carrier concentration distribution of electrons in the conduction band under dark conditions (broken lines, $n_p$ and $n_n$) under illumination condition (solid lines, $n'_p$ and $n'_n$).
Carrier concentrations are shown for three conditions: $V = 0$ (the short-circuit condition), $V_{OC}$ (the open-circuit condition), and $V = V^*$ (the crossover point), respectively.
The carrier concentrations far from the depletion layer are given by: $n_{p0}' = n_{p0} + \Delta n_p$, $n_{n0}' = n_{n0} + \Delta n_n$, and the excess carrier concentrations are given by: $\Delta n_p = G \tau_n$, $\Delta n_n = G \tau_p$. } 
\label{fig2}
\end{figure}

Here, we revisit the classical one-dimensional p-n junction model illustrated in Figure \ref{fig2} (a). 
The model consists of p-type and n-type regions, and the width of the depletion layer $W$ is much smaller than the diffusion lengths of minority carriers in both regions. 
Within the simplest p-n junction model, we show that the photocurrent equation, Eq. (\ref{I_ph2}), can be derived analytically.
It indicates that the model already contains the essential physics governing the photocurrent as a function of the applied voltage and built-in potential; however, this aspect has received little attention in previous studies.  

The electron carrier concentration in the p-type region, denoted by $n_p$, is governed by the following drift-diffusion equation with respect to position $x$ and time $t$:
\begin{eqnarray}
\frac{\partial n_p}{\partial t} = D_n\frac{\partial ^2n_p}{\partial x^2} +\mu_n\frac{\partial n_pE}{\partial x}- \frac{n_p-n_{p0}}{\tau_n} +G.
\label{n_p_eq}
\end{eqnarray}
Here, the first and second terms represent the spatial derivatives of the diffusion current and the drift current, respectively.
$D_n$, $\tau_n$, $\mu_n$ and $E$ are the diffusion coefficient, the minority carrier lifetime for electrons, the mobility of electrons and electric field, respectively
$n_{p0}$ denotes the carrier concentration under thermal equilibrium conditions. 
In this manuscript, $G$ is assumed to be uniform throughout the solar cell for simplicity.


As well as $n_p$, the carrier concentration of holes in n-type region, $p_n$, is expressed by 
\begin{eqnarray} 
\frac{\partial p_n}{\partial t} = D_p\frac{\partial ^2p_n}{\partial x^2} -\mu_p\frac{\partial p_nE}{\partial x}- \frac{p_n-p_{n0}}{\tau_p} +G
\label{p_n_eq}
\end{eqnarray}
$D_p$, $\tau_p$, $\mu_p$ and $p_{n0}$ are the diffusion constant, minority carrier lifetime, mobility, and the thermal equilibrium concentration for the holes, respectively.


In the steady state, time derivative is zero, such as $\frac{\partial n_p}{\partial t} = 0$ and $\frac{\partial p_n}{\partial t} = 0$.
For analytical tractability, the drift current of minority carriers is typically neglected, as the electric field $E$ is assumed to be negligible in both the p-type and n-type quasi-neutral regions.
As a first approximation, we assume that the diffusion currents in the p- and n-regions dominate over the drift current in the depletion region.
Hence, we obtain the following derivative equations to solve;
\begin{eqnarray} 
L_n^2\frac{d^2n_p}{dx^2} = n_p-\left\{n_{p0} +G\tau_n\right\}\ \ \mathrm{for}\ (x \leq x_p),
\label{n_p_eq1}
\end{eqnarray}
and 
\begin{eqnarray} 
L_p^2\frac{d^2p_n}{dx^2} =p_n-\left\{p_{n0} +G\tau_p\right\}\ \ \mathrm{for}\ (x \geq x_n).
\label{p_n_eq1}
\end{eqnarray} 
Here, $L_n = \sqrt{D_n\tau_n}$ and $L_p = \sqrt{D_p\tau_p}$ are the minority carrier diffusion lengths.
The solutions are obtained as follows,
\begin{eqnarray} 
n_p(x) &=& N_p \exp\left(\frac{x-x_p}{L_n}\right) + n_{p0} +G\tau_n\ \ \mathrm{for}\ (x \leq x_p),
\label{n_p}
\end{eqnarray}
and
\begin{eqnarray} 
p_n(x) &=& P_n \exp\left(-\frac{x-x_n}{L_p}\right) + p_{n0} +G\tau_p\ \ \mathrm{for}\ (x \geq x_n).
\label{p_n}
\end{eqnarray}
$x_n$ and $x_p$ are the edge positions of p-type and n-type regions, respectively, as shown in Figure \ref{fig2} (b).
$N_p$ and $P_n$ are constants of integration.
At the position far from the depletion layer, the minority carrier concentrations are given by $n_{p0} +G\tau_n\ \ \mathrm{for}\ (x \leq x_p)$ and $p_{n0} +G\tau_p\ \ \mathrm{for}\ (x \geq x_n)$.

Within the depletion region, carrier recombination and generation are typically assumed to be negligible. 
Based on the Boltzmann distribution, the boundary conditions across the depletion region are expressed as follows:
\begin{eqnarray} 
n_p(x_p)&=& n_n(x_n) \exp\left\{-\frac{q(V_{bi}-V)}{k_BT}\right\},
\label{BoltzmannDistribution}
\end{eqnarray}
and
\begin{eqnarray} 
p_n(x_n) &=& p_p(x_p) \exp\left\{-\frac{q(V_{bi}-V)}{k_BT}\right\}, 
\label{boundary}
\end{eqnarray}
respectively.

We are now at a turning point, leading to either Eq. (\ref{I_ph}) or Eq. (\ref{I_ph2}).
If the concentrations of photoexcited carriers are assumed to be negligible compared to the majority carrier concentrations in thermal equilibrium, the majority carrier concentrations are given by
$p_p(x) \sim p_{p0} \ \ \mathrm{for}\ (x \leq x_p)$
in the p-type region and
$n_n(x) \sim n_{n0} \ \ \mathrm{for}\ (x \geq x_n)$
in the n-type region.
Using these relations, Eq. (\ref{I_ph}) is derived.

However, it is more accurate to take into account the photoexcited majority carriers under illumination. 
Charge neutrality requires that the excess majority-carrier density equals the excess minority-carrier density.
The concentration of photoexcited majority carriers is equal to that of photoexcited minority carriers, since photoexcitation always generates electron-hole pairs.
The excess carrier concentrations are given by $\Delta n_p = G \tau_n = \Delta p_p$  in the p-type region, and $\Delta p_n = G \tau_p = \Delta n_n$ in the n-type region, respectively.
Therefore, the majority carrier concentrations can be expressed as follows:
\begin{eqnarray} 
p_p(x) &=&  p_{p0} + \Delta n_p = p_{p0} + G\tau_n\ \ \mathrm{for}\ (x \leq x_p).
\label{pp}
\end{eqnarray}
in the p-type region and
\begin{eqnarray} 
n_n(x) &=&  n_{n0} + \Delta p_n =n_{n0} + G\tau_p\ \ \mathrm{for}\ (x \geq x_n),
\label{nn}
\end{eqnarray}
in the n-type region. 
Here, it is assumed that the diffusion current is negligibly smaller than the drift current in the majority carrier region, because the majority-carrier concentrations are much larger than the minority carrier concentrations.

By applying the boundary conditions given in Eqs. (\ref{pp}) and (\ref{nn}), and the thermal equilibrium conditions between the p-n junction, $n_{p0} =n_{n0} \exp\left(-\frac{qV_{bi}}{k_BT}\right)$ and $p_{n0} =p_{p0} \exp\left(-\frac{qV_{bi}}{k_BT}\right)$, 
the constants of integration, $N_p$ and $P_n$, are obtained by:
\begin{eqnarray} 
N_p = n_{p0}\left\{\exp\left(\frac{qV}{k_BT}\right) - 1\right\} -G\tau_n\left\{1 - \frac{\tau_p}{\tau_n}\exp\left(-\frac{q(V_{bi}-V)}{k_BT}\right)\right\}, 
\label{NP2}
\end{eqnarray}
and
\begin{eqnarray} 
P_n= p_{n0}\left\{\exp\left(\frac{qV}{k_BT}\right) - 1\right\} -G\tau_p\left\{1 -\frac{\tau_n}{\tau_p}\exp\left(-\frac{q(V_{bi}-V)}{k_BT}\right)\right\}. 
\label{PN2}
\end{eqnarray}
The current flowing across the p-n junction consists of the diffusion currents of electrons and holes at the edges of the depletion layer.
The electron current $I_n$ and the hole current $I_p$ are given by:
\begin{eqnarray} 
I_n &=& +qAD_n\frac{d n_p}{dx}\Bigm\vert_{x=x_p} = \frac{qAD_nN_p}{L_n},
\end{eqnarray}
and
\begin{eqnarray} 
I_p &=& - qA D_p\frac{d p_n}{dx}\Bigm\vert_{x=x_n} = \frac{qAD_pP_n}{L_p}.
\end{eqnarray}
Then, finally total current $I = I_n + I_p$ is obtained by
\begin{eqnarray} 
I &=& I_s\left\{\exp\left(\frac{qV}{k_BT}\right)-1\right\}-qA G (L_n + L_p)\left\{1-\alpha\exp\left(-\frac{q(V_{bi}-V)}{k_BT}\right)\right\},
\label{Ifinal}
\end{eqnarray}
where $I_s = qA\left(\frac{D_n n_{p0}}{L_n} + \frac{D_p p_{n0}}{L_p}\right)$ is the reverse saturation current, and
\begin{eqnarray} 
\alpha = \frac{L_n\tau_p^2+L_p\tau_n^2}{(L_n+L_p)\tau_p\tau_n} 
\end{eqnarray}
is a positive constant parameter.
The first term in Eq. (\ref{Ifinal}) represents Shockley's diode equation, which describes the current under dark conditions \cite{Shockley1949}.
The second term represents the photocurrent generated by the photovoltaic effect, which corresponds to Eq. (\ref{I_ph2}).
We note that voltage-dependent expression of photocurrent similar to Eq. (\ref{I_ph2}) was proposed for organic polymer solar cells \cite{Gaur2013PinPhoto,Kumar2013JAP}. 
For thin film solar cells,  similar expression has been derived from a p-i-n junction model \cite{Moore2014IEEE}.
However, such voltage-dependent photocurrent models have not been established for crystalline silicon p-n junction solar cells.

In crystalline silicon solar cells, the built-in potential $V_{bi}$ is typically designed to be significantly greater than the thermal voltage $k_B T/q$ at room temperature.
As a result, the difference between  Eq. (\ref{I_ph}) and Eq. (\ref{I_ph2})  becomes negligibly small. 
Eq. (\ref{Ifinal}) yields the following expression of the short-circuit current;  
\begin{eqnarray}
I_{SC} =-qA G (L_n + L_p)\left\{1 - \alpha \exp\left(-\frac{qV_{bi}}{k_B T}\right)\right\},
\label{I_SC}
\end{eqnarray}
and for the open-circuit voltage; 
\begin{eqnarray} 
V_{OC} =\frac{k_BT}{q}\ln\frac{I_s+\alpha qA G (L_n + L_p)} {I_s+\alpha qA G (L_n + L_p)\exp\left(-\frac{qV_{bi}}{k_B T}\right)}.
\label{V_OC}
\end{eqnarray}
The differences in $I_{SC}$ and  $V_{OC}$  predicted by the conventional  and present expressions are negligible when $qV_{bi}/k_B T \gg 1$ . 
Consequently, the voltage dependence of the photocurrent in crystalline silicon solar cells has largely remained unnoticed.

This equation indicates that the photocurrent consists of the conventional component, $-qA G (L_n + L_p)$, and a backward component that depends on the built-in potential $V_{bi}$, and the applied voltage $V$, given by $ qA G (L_n + L_p)\exp\left\{-\frac{q(V_{bi} - V)}{k_B T}\right\}$.
The backward photocurrent is essential for a complete description of the photovoltaic effect in p-n junctions.
Since two current components flow in opposite directions, a voltage exists at which they become equal in magnitude. 
This leads to the appearance of a crossover point in the $I-V$ characteristics.
The crossover point $(V^*,\ I^*)$ in the $I-V$ characteristics is given by:
\begin{eqnarray} 
V^* = V_{bi} - \frac{k_BT}{q}\ln\alpha,
\label{V*}
\end{eqnarray}
and 
\begin{eqnarray} 
I^* = I_s\left\{\frac{1}{\alpha}\exp\left(\frac{qV_{bi}}{k_BT}\right)-1\right\}.
\label{I*}
\end{eqnarray}
Since $V^*$ and $I^*$ are generally positive, the crossover point typically lies in the first quadrant of the $I-V$ characteristics, as shown in Figure \ref{fig1}.
It is noteworthy that both $V^*$ and $I^*$ are independent of $G$, therefore, the crossover point is a fixed point under any illumination intensity.
$V^*$ does not equal $V_{bi}$ when parameters in p- and n- regions are asymmetric.  
Except for the special case of perfectly symmetric p- and n-regions, $V^*$ does not coincide with $V_{bi}$ due to $\alpha \neq 1$.

Figures \ref{fig2} (c)-(e) illustrate the electron concentration distribution in the conduction band for $V = 0$, $V = V_{OC}$, and $V = V^*$, respectively.
The spatial derivative of the carrier concentration at $x_p$ determines the diffusion current of electrons across the p-n junction.
Figure \ref{fig2} (c) illustrates the short-circuit condition, where $dn_p/dx|_{x=x_p} =0$ under dark conditions, and $dn'_p/dx|_{x=x_p} \neq 0$ under illumination. 
Figure \ref{fig2} (d) illustrates the open-circuit condition in which $dn_p /dx|_{x=x_p} \neq 0$ and $dn'_p/dx|_{x=x_p} = 0$.
At the crossover point shown in Figure \ref{fig2} (e), the spatial derivatives of the electron concentration under dark and illuminated conditions become equal at $x_p$, such that $dn_p/dx|_{x=x_p} = dn'_p/dx|_{x=x_p}$. 
Since the diffusion current is therefore identical in both cases, the resulting photocurrent of electrons becomes zero.
A similar situation applies to holes in the valence band.
Even if the crossover points for electrons and holes differ, there must exist a unique point at which the total current becomes independent of the excess carrier concentrations.

This theoretical analysis indicates that the photocurrent inherently consists of two components: the well-known conventional constant photocurrent and a voltage-dependent backward photocurrent. 
Under illumination, incident photons excite both minority and majority carriers.
The photoexcited majority carriers diffuse into minority carrier regions, while the photoexcited minority carriers diffuse into majority carrier regions.
The former constitutes the conventional photocurrent, whereas the latter corresponds to the backward photocurrent.
The built-in potential serves to suppress the diffusion of photoexcited majority carriers across the p-n junction, while leaving the diffusion of photoexcited minority carriers unaffected.
This asymmetry is essential for charge separation.




\section{Experimental}
We experimentally demonstrate the crossover point between the $I-V$ curves under dark and illuminated conditions using lock-in photovoltaic measurements on a commercial monocrystalline silicon solar cell.
The solar cell used in the experiment was the KXOB25-12X1F model (ANYSOLAR) with a light absorption area of 22 mm $\times$ 7 mm.
The presence of a single junction in the cell enabled clear observation of the intrinsic characteristics of the p-n junction.
Illumination was provided by a basic solar simulator, XC-100EFSS (SERIC), providing a central irradiance of 1000 $\mathrm{W/m^2}$ at a distance of 38 cm.

There was a crucial problem in directly comparing $I-V$ curves under dark and illuminated conditions, as the cell temperature varies with light intensity.
Since the $I-V$ characteristics of solar cells vary with temperature, the crossover point could potentially appear as an artifact of temperature effects.
The temperature increase at the solar cell surface caused by light irradiation could not be precisely controlled using an electric heater.
Therefore, we employed a lock-in detection technique to accurately measure the small voltage difference between dark and illuminated conditions, while keeping the cell temperature constant.

\begin{figure}[t]
\centering
\includegraphics[width=0.8\linewidth]{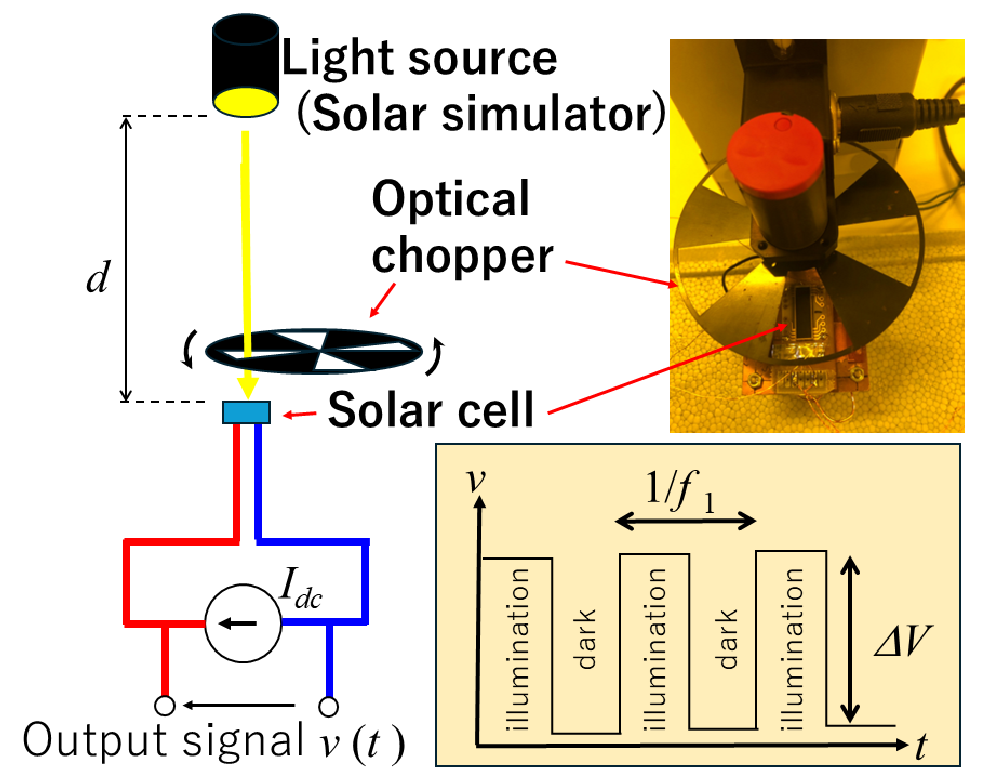} 
\vspace{0cm}
\caption{Experimental setup for lock-in photovoltage measurement of a single p-n junction solar cell.
An optical chopper rotates above the solar cell device, periodically shading light from the solar simulator at a frequency of $f_1$.
A constant DC current, $I_{dc}$, is applied to the solar cell using a DC current source (Keithley 6220).
The output voltage between the electrodes of the solar cell exhibits a periodic signal with a fundamental frequency of $f_1$.
The voltage difference $\Delta V$ represents the difference between the voltages under dark and illuminated conditions.
The fundamental Fourier component, $V_1$ of the Fourier serise expansion, $v(t)= V_0 + V_1 \cos(2\pi f_1 t + \phi_1) + V_2 \cos(4 \pi f_1 t+\phi_2) +\cdots$, is proportional to $\Delta V$.
Therefore, $V_1$ is measured using a lock-in amplifier synchronized to the reference frequency $f_1$. }
\label{fig3}
\end{figure}

A rotary shutter positioned above the cell periodically interrupted the light exposure at a shutter frequency $f_1$, as shown in Figure \ref{fig3}.
As a result, the cell voltage $v(t)$ exhibited periodic variation at the same frequency $f_1$.
Using a lock-in amplifier, the fundamental component $V_1$ of the oscillating signal $v(t)$ was measured. 
The voltage difference $\Delta V$ (illustrated in Figure \ref{fig3}) between the dark and illuminated conditions is proportional to $V_1$, and depends on the applied DC current.
When the shutter period $1/f_1$ is sufficiently shorter than the time constant of the solar cell’s temperature change, the temperature can be regarded as constant.

\begin{figure}[t]
\centering
\includegraphics[width=0.8\linewidth]{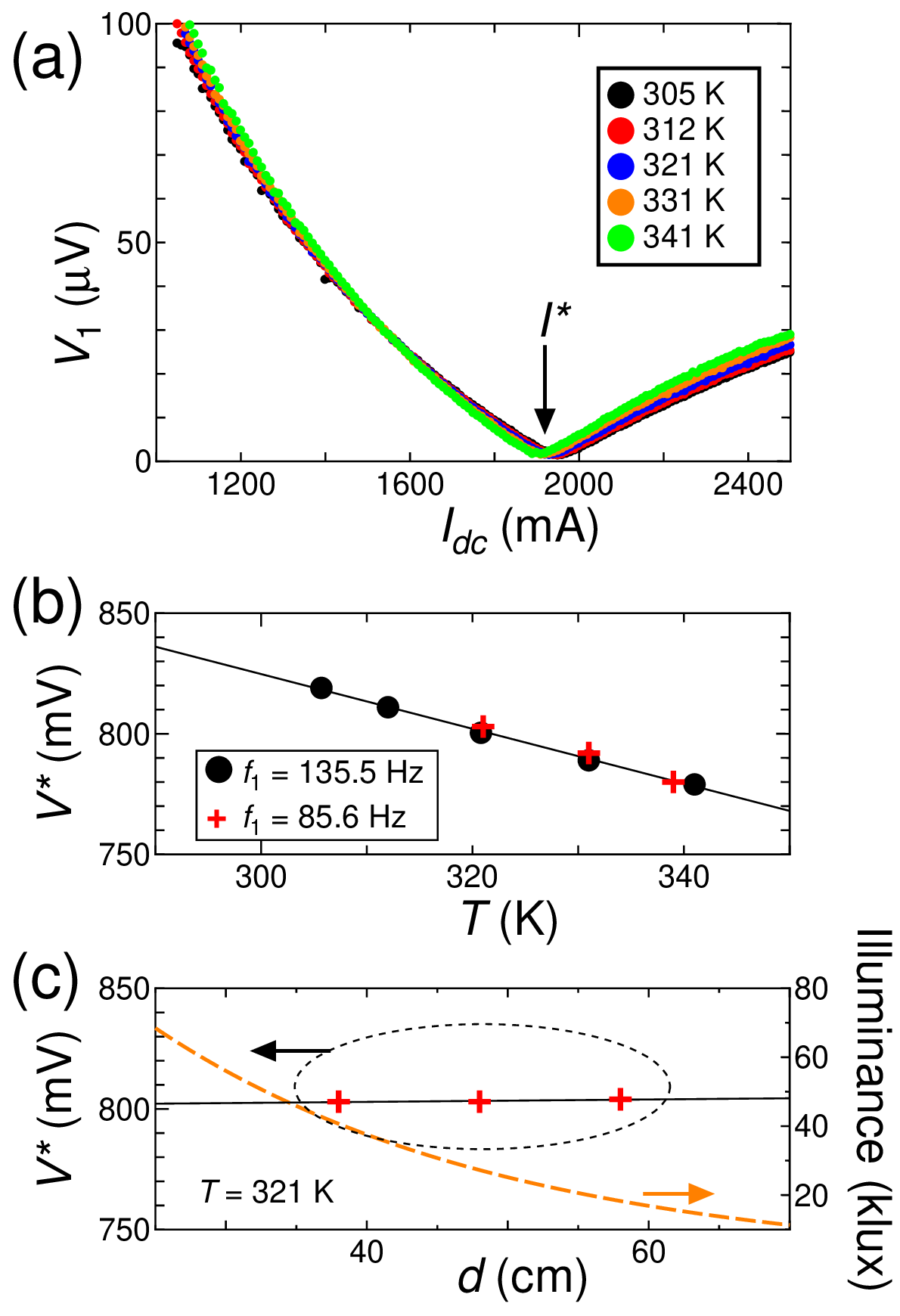} 
\vspace{0cm}
\caption{(a) Fundamental Fourier component $V_1$ of output signal $v(t)$ as a function of applied DC current measured using the lock-in technique at various temperatures.
The frequency of the optical chopper is $f_1 = 135.5\ \mathrm{Hz}$.
(b) Temperature dependence of the crossover voltages for different chopper frequencies.
(c) Crossover voltages under different illuminance conditions, measured with $f_1 = 85.6\ \mathrm{Hz}$.
$d$ is the distance between the light source and the solar cell, as depicted in Figure \ref{fig3}.
}
\label{fig4}
\end{figure}

Figure \ref{fig4} (a) shows the experimental results of the lock-in measurements conducted at several temperatures. 
The temperature is measured using a thermoresistance sensor attached to a copper block with heater control. 
This block is thermally coupled with the solar cell, which is also equipped with its own heater control.
The fundamental voltage $V_1$ at the chopper frequency ($f_1 = 135.5 \ \mathrm{Hz}$) varies with the applied DC current. 
Within the range of 1900-2000 mA, $V_1$ reaches a minimum. 
The minimum corresponds to the crossover point.
Because the lock-in amplifier measures only the signal amplitude, $V_1 > 0$ is always positive, causing the curve to fold back above the minimum.

The current at the minimum corresponds to  $I^*$, and the corresponding voltage immediately gives $V^*$ .
A typical value of  $V^*$  is around $800\ \mathrm{mV}$, which is about three orders of magnitude larger than $V_1$ or $\Delta V$. 
Therefore, it is difficult to observe this crossover point using conventional I–V measurements.
This may be one reason why the crossover phenomenon has received little attention in crystalline silicon solar cells.

Figure \ref{fig4} (b) shows the crossover point $V^*$ as a function of cell temperature.
$V^*$  was found to be independent of the chopper frequency, indicating that the crossover is not caused by light-induced heating.
The values of $V^*$ decreased as the cell temperature increased.
This behavior closely resembles that of the built-in potential in silicon diodes, as characterized by alternative measurement techniques\cite{Ikram2009EJP,Garland2011SolarEnergy}.
Therefore, the experimental observation of the temperature dependence of $V^*$ appears to be consistent with Eq. (\ref{V*}) when the temperature dependence of $\ln\alpha$ is negligible.  

Figure \ref{fig4} (c) illustrates the crossover voltage, $V^*$, as a function of the distance from the light source, $d$. 
Although the illuminance on the solar cell varied with $d$,  the crossover voltage remained unchanged.
Since the optical generation rate $G$ is proportional to the number of incident photons with energy above the bandgap, $G$ increases with light intensity.
Nevertheless, the experimental $V^*$ values appear to be independent of intensity.
This observation is consistent with the prediction of Eq. (\ref{V*}).
These experimental results provide strong evidence for the presence of a voltage-dependent backward photocurrent and support the validity of the improved analytical solution for the photocurrent.

Realistic  $I-V$  characteristics depend on a variety of factors, such as the ideality factor, series and shunt resistances, detailed doping profiles, and band-structure effects. 
The present expression can be generalized to include these parameters, allowing quantitative agreement with experimental  $I-V$ characteristics.
For example, by introducing the series and shunt resistances, $R_s$ and $R_p$, respectively, Eq. (\ref{Ifinal}) can be rewritten as
\begin{eqnarray} 
I = I_s\left\{\exp\left(\frac{q(V - R_s I ) }{k_BT}\right)-1\right\}-qA G (L_n + L_p)\left\{1-\alpha\exp\left(-\frac{q(V_{bi}-V + R_s I) }{k_BT}\right)\right\}+\frac{V - R_s I }{R_p}.
\label{Ifinal2}
\end{eqnarray}
Even in this form, there must be a unique crossover point that is independent of $G$.
Therefore, the illumination-intensity independence of the crossover voltage, $V^*$, indicates the intrinsic nature of the photovoltaic effect.

\section{Conclusion}
We have shown that the photocurrent in monocrystalline silicon solar cells is not strictly constant but exhibits a voltage dependence arising from the built-in potential. 
As a consequence, even conventional crystalline silicon solar cells are predicted to exhibit crossover behavior in the $I-V$ characteristics between dark and illuminated conditions.
This result revisits a fundamental issue in the classical theory of the photovoltaic effect in p-n junctions, namely the quantitative role of the built-in potential in photocurrent generation. 
We derived an analytical expression for the photocurrent that explicitly incorporates the built-in potential without introducing specific device-dependent parameters.
The proposed expression predicts the existence of an illumination-independent crossover point in the $I-V$ characteristics.
This prediction was experimentally verified in a commercial monocrystalline silicon solar cell using an optical lock-in measurement technique.
The present formulation clarifies why the influence of the built-in potential has remained largely hidden in conventional descriptions of crystalline silicon solar cells and provides a more complete understanding of its role in photovoltaic operation.

\section*{Acknowledgements}
We are grateful for the financial support provided by the Yashima Environment Technology Foundation. 
We also appreciate the budgetary support for promoting research activities at NIT, Fukui College.


\newpage

\end{document}